\documentstyle[preprint,aps,psfig]{revtex}
\tightenlines
\input psfig.sty

\begin{document}
\draft

\preprint{IU/NTC-01-10}
\title{Small $x$ divergences in the Similarity RG approach to LF QCD}
\author{Martina Brisudova\thanks{{\it On leave from Physics Inst., SAS, 
Bratislava.} Electronic address: {\tt brisuda@phys.ufl.edu}}}
\address{ Dept. of Physics, University of Florida,
 P.O. Box 118440, Gainesville, FL 32611 \\}
\address{ {\it and} \\}
\address{Nuclear Theory Center, Indiana University, 2401 Milo B. Sampson
 Lane,
 Bloomington, IN 47408}
\date{November 15, 2001}
\maketitle

\begin{abstract}
We study  small $x$ divergences in  boost invariant similarity
 renormalization group 
 approach to light-front  QCD in a heavy quark-antiquark state. With the
 boost
 invariance maintained, the infrared divergences do not cancel out in
 the physical states,
 contrary to previous studies where boost invariance was violated by a
 choice of a renormalization scale. This
 may be an indication that the zero mode, or nontrivial light-cone
 vacuum structure, might
 be important for recovering full Lorentz  invariance.
\end{abstract}

%\pacs{ }

\section{Introduction}
 Light-cone or light-front \cite{dirac} description of field theories
 has long 
attracted attention 
for various reasons, often
related to its peculiar kinematics \cite{coester}. Light-cone
 coordinates are 
used in vastly different applications usually in Hamiltonian formulation
 (see 
\cite{thelongpaper} for an extensive list of references), ranging from
 deep 
inelastic scattering and phenomenology
to attempts to formulate the theory of everything.
Our own interest in light-front field theories is motivated by an
 assumption 
that in light-front coordinates it is possible to derive from first
 principles 
 a self-consistent and systematic constituent approximation to QCD  
\cite{thelongpaper}.  We will give more details below.
 
As a  
consequence of the light-front kinematics, one component of the 
three-momentum, 
so called longitudinal momentum $k^+$, is positively definite and can be
interpreted as playing role of a Newtonian mass \cite{weinberg,soskel}.
 Thorn 
has long advocated that for this reason, light-front formulation of
 string 
theory is one of the best hopes for a truly fundamental,
 non-perturbative 
description of strings \cite{thorn} and to quantitatively realize the 
conjectured correspondence between string and field theory 
\cite{zasetentrn}.

 Because of the above mentioned properties of the longitudinal momentum
 $k^+$, 
light-front vacuum was thought to be trivial. This notion turned out to
 be 
naive, but the light-front vacuum indeed can contain only particles with
$k^+=0$ known as the zero mode. Therefore, the light-front vacuum {\it
 can be made} trivial by imposing appropriate cutoffs. 
This is the last point that the light-front community agrees on. 
Opinions differ when it comes to the question how to proceed further.

Basically, there are two approaches to the problem of light-front vacuum
 \cite{dave}. One 
is to address it head-on \cite{glazek0}. Typically, this is done for
 various, often lower
 dimensional, field theories (for 
review see \cite{BrodskyPauliPinsky}, for extensive list of references
 see 
also \cite{thelongpaper}) in the context of discreet 
light cone quantization (DLCQ) (for a recent review of the method as
 well as an expansive list of references see \cite{DLCQ}). The
theory is formulated in a box of a finite longitudinal size which makes
 $k^+$ 
discreet and non-zero. The zero mode \cite{zeromode} is then determined
 as a solution to constraint 
equations. Constraint and dynamical zero modes are often discussed.
 Critics complain that so far, it 
is not clear how much of the zero mode determined in this manner is an 
artifact of the method (in other words, a 
counterterm specific to the DLCQ) and whether the continuum limit
 exists\footnote{See for
example, transcript of discussion sections \cite{wilsonja} at the Fourth
Workshop on Light-Front Quantization and Non-Perturbative Physics.}.

As an untypical example of direct studies of the zero mode, we would
 like to mention a 
recent work by Tomaras, Tsamis and Woodard \cite{woodard} on 
back reaction in light-cone QED. Though motivated by the back-reaction
 in quantum 
gravity occurring on an inflating background, their work addresses some
 issues 
of the light-front vacuum without having to evoke DLCQ. They had
 constructed 
a full operator solution to a free QED  coupled to a constant external
 electric field 
in continuum (3+1) dimensions. In this set up, all modes are forced to
 go through the zero 
mode at which point particle pairs are created. The zero mode of the 
constraint components of the fermionic field is shown to be crucial for 
unitarity.

The other approach to the problem of zero mode is more pragmatic.
 Instead of 
trying to solve for the zero mode, it is simply cutoff, be it with DLCQ 
\cite{joelthorn} or an explicit infrared cutoff in a continuous
 formulation 
\cite{thelongpaper}. Physics associated with this mode can then be put
 in 
form of counterterms, if needed, for example, to restore symmetries or
account for phenomena associated with the vacuum. Traditionally,
 spontaneous
 symmetry breaking was viewed as an example of such a phenomenon 
\cite{thelongpaper,zeromode,wilsonja}. However, Rozowsky and Thorn
 \cite{joelthorn} 
have argued recently that, while conceding that the inclusion of a 
fundamental zero mode is a valid 
theoretical option, it is not necessary to describe spontaneous symmetry
breaking where its presence seems to be most needed.  Indeed, in scalar 
quantum field theory in (1+1) dimensions DLCQ the physics of spontaneous
 symmetry breaking is completely and accurately described without the
 zero modes
 \cite{joelthorn}. 

The need to put in new counterterms associated with the infrared (IR)
 regulator in our continuum 
formulation \cite{thelongpaper} was 
anticipated \cite{wilsonja} but was not  encountered yet in the
applications to hadronic physics so far \cite{us}. Perry \cite{P2} has
 shown that even
though the one body and two-body effective operators are each separately
 divergent as
$k^+$ goes to zero, the divergences {\it exactly} cancel in any color
 singlet 
state. The cancelation does not occur for non-singlet states, leaving
 them with an infinite mass. This together with a naturally generated
confining potential (imprecisely referred to as ``logarithmic'') is a
 plausible
feature of the approach. Note that both the effective confining
 potential and the 
infinite mass of the color non-singlets originate from small $k^+$
 regions.
  
A shortcoming of the above described result is that it was obtained in a
similarity renormalization group limited to matrix elements that
required a specific infrared regulator (theta function) and an
 introduction of an
arbitrary scale ${\cal P}^+$ which violates an explicit kinematic
 symmetry of light
cone. In a bound state calculation one can argue that there is a
 preferred scale,
 i.e. one associated with the typical longitudinal momentum of the
 state, or,
the total center of mass $P^+$; however, consequences of 
the violation of the kinematic symmetry  are not known.

Since then, the similarity renormalization group approach has advanced
 so that 
it is no longer necessary to violate the kinematic boost invariance, and
 to generate counterterms dependent on the total center of mass $P^+$. 
 It is also possible to use an arbitrary form of 
the small $x$ regulator \cite{glazkovia} ($x$ being a dimensionless,
 boost invariant fraction
of the total $P^+$). Thus, for the first time we are able to study with
 some degree of
generality the issues related to the light-cone zero mode. 
Recently, G{\l}azek has found \cite{gluons}
that even though infrared divergent terms cancel out in the running
 coupling, there is a residual finite dependence on the functional form 
of the infrared  regulator.  In this paper, we wish to study 
the issue of small $x$ (or infrared) divergences in color singlet states
consisting of a heavy quark and antiquark of the same flavor, for
 simplicity.

The paper is organized as follows.
Section II is devoted to an overview of the similarity renormalization
 group approach light-front QCD. We start with a general description of 
the approach.  For the sake of making 
this paper self-contained,
the reader is reminded of the structure of the classical light-front
 Hamiltonian. The corresponding quantum Hamiltonian cannot be defined
 without
regularization. Details of regularization determine structure and form 
of counterterms. In this  case, details of regularization 
are particularly important, because only one of the regulators can be
properly removed by renormalization. After introducing regularization,
 we
present a technical description of the boost invariant similarity 
renormalization group for  particles. Remaining
sections deal specifically with quark-antiquark color singlets to second
 order in $g$.
Section III contains the effective one body operators to second order in
$g$; section IV the effective two body operators to the same order. In
 section V we address
the issue of infrared divergences in those operators. We conclude with a
 short summary and
conclusions.

\section{Similarity renormalization group approach to QCD}

In this section we  briefly review the similarity renormalization group 
 (RG) approach to
 light-front (LF) QCD, introduced in ref. \cite{thelongpaper}. We show
 the unregulated
 canonical light-front Hamiltonian. Then the regularization that we use
 is briefly explained,
 and finally, the similarity renormalization procedure is outlined.
 
 The basic assumption behind the approach is that it is possible {\it to
 derive} a constituent
 picture of hadrons from QCD. To separate  vacuum fluctuations, it is
 convenient to use
 light-cone coordinates with cutoffs preventing zero longitudinal
 momentum. Then 
 vacuum is forced to be trivial. In addition, since such a cutoff
 introduces a nonzero
 minimum for the (kinematic, positive) longitudinal momentum,  it also
 restricts number
 of particles in any state with a fixed longitudinal momentum $P^+$. For
 massive particles,
 due to light-cone free energy {\it increasing}  with decreasing
 longitudinal momentum, 
 many-body states tend to have higher free energy than few body states.
 Mixing
 of low energy few particle states with many-body states can be expected
 to be naturally
 small, at least at weak coupling. These features make for useful
 prerequisites toward the
 constituent picture of hadrons.
 
 The apparent difficulty with renormalization of light-front
 Hamiltonians (compared to
Lagrangians) is turned into an advantage by using  similarity
 renormalization 
 \cite{similarity}. The basic
 idea of the similarity renormalization group is simple. The regulated
 bare Hamiltonian that
  mixes all energy scales is transformed via a unitary transformation to
 a Hamiltonian
 that contains direct couplings only between neighboring scales. Such a
 Hamiltonian is
  referred to  as "band-diagonal" in the sense that it does
 not allow for direct couplings between arbitrary scales. 
 At any finite order of perturbation theory, a band-diagonal Hamiltonian
 cannot produce any
ultraviolet divergences providing its matrix elements are finite.
 Therefore, by requiring that
the band-diagonal Hamiltonian be independent of the regulator,
 counterterms that need to
be added to the bare Hamiltonian can be identified.

Upon completion  of renormalization, the effective band diagonal
 Hamiltonian is subject to diagonalization. We refer the reader to
 \cite{us} for more details regarding this step of the procedure.

\subsection{Classical light-front QCD Hamiltonian}
The starting point is  the canonical light-front QCD Hamiltonian in
 light-cone gauge,
 $A^+_a=0$. We
will not explicitly show terms that are not important for the specific
calculations presented in the next sections.
 For a detailed discussion of the light-front
Hamiltonian see  \cite{thelongpaper,xy}. Purely gluonic terms are
 studied in\cite{gluons}. 

The part of the classical {\it unregulated} 
canonical Hamiltonian that is relevant for our study of  
the bound state of a quark and antiquark  of the same flavor is, 
\begin{eqnarray}
H= H_{\rm free} + V_1 +V_2   \   \  ;
\end{eqnarray}
where $H_{\rm free}$ is the free light-front Hamiltonian:
\begin{eqnarray}
H_{free} & = & \int dx^- d^2 x_{\perp} \left( {1\over{2}} \bar{\psi}
 \gamma^+
{-{\partial}^{\perp 2}+m^2\over{i{\partial}^+}} \psi - {1\over{4}}
 (\partial^{\mu} A^{\nu} -\partial^{\nu} A^{\mu})
(\partial_{\mu} A_{\nu} -\partial_{\nu} A_{\mu}) \right)
,
\end{eqnarray}
\noindent and 
\begin{eqnarray}
V_1= g \int dx^- d^2 x_{\perp} \bar{\psi}
{\not{\hbox{\kern-4pt $A$}}} \psi
\end{eqnarray}
contains the standard order $g$ quark-gluon coupling. Here $\psi$ and
 $A^{\mu} \equiv \sum_{a} A_a^{\mu} T^a$ are free
light-front fields:
\begin{eqnarray}
\psi =\left( \begin{array}{c} \psi_+ \\
\psi_- =
{1\over{\partial ^+}}(-i\vec{\alpha}^{\perp}\cdot \vec{\partial}_{\perp}
+\beta m) \psi_+ \end{array} \right)
\end{eqnarray}
and
\begin{eqnarray}
A^{\mu} = \left( A^+=0,\  A^-={2\over{\partial
 ^+}}\vec{\partial}^{\perp} \cdot
\vec{A}^{\perp}, \  \vec{A}^{\perp} \right).
\end{eqnarray}
The constrained fields contain, in addition to $\psi _-$ and $A^-$ shown
 above, ${\cal O}(g)$ terms that are fully determined
functions of the physical degrees of freedom. These produce new terms in
the canonical Hamiltonian, among which
\begin{eqnarray}
V_2 =  - 2g^2 \int dx^- d^2 x_\perp
 (\psi_+^{\dagger} T^a \psi_+ ) \left( \frac{1}
{\partial^+} \right)^2 (\psi_+^{\dagger} T^a \psi_+)
\end{eqnarray}
is the so-called instantaneous gluon exchange between two fermions.
As the next step,  fields are expanded in mode functions consisting of
 (light-cone)
spinors $u_{p \sigma}, v_{p \sigma}$ \cite{brodskylepejdz}, ($p$ denotes
 momentum,
$\sigma$ denotes spin),
SU(3) color spinors $\chi_c$ and plane waves,
\begin{eqnarray}
\psi \left( x^-, x^{\perp}\right) = \sum_{{c \sigma}}\int
 {d^3p\over{(2\pi)^3 2 p^+}}\left[
\chi_c u_{p \sigma} e^{-i \, p \,  x} \ b_{p, \sigma , c } + \chi_c
 v_{{p \sigma}} e^{i \, p \,  x}
d_{p, \sigma , c }^{\dagger} \right] .
\end{eqnarray}
The Hamiltonian is expressed in terms of what will become, after
 quantization,
 creation and annihilation
 operators of fermions ($b^{\dagger}$,  $d^{\dagger}$, $b$, $d$) and 
 gluons $a^{\dagger}$, $a$.
 For example, the kinetic part $H_{free}$ becomes 
\begin{eqnarray}
H_{free} & = & 
\sum_{c \sigma} \int {d^3p\over{(2\pi)^3 2 p^+}} {p^{\perp
 2}+m^2\over{p^+}}
\left( b_{p c \sigma}^{\dagger}b_{p c \sigma} + d_{p c
 \sigma}^{\dagger}d_{p c \sigma}\right) 
 + \sum_{c \sigma} \int {d^3k\over{(2\pi)^3 2 k^+}} {k^{\perp
 2}\over{k^+}}
 a_{p c \sigma}^{\dagger}a_{p c \sigma}\  \  \label{h0}.
\end{eqnarray}
where $E_{p} \equiv {p^{\perp 2}+m^2\over{p^+}}$ 
and $E_{k} \equiv {k^{\perp 2}\over{k^+}}$ are free light-front energies
for massive and massless particle, respectively.

As another example, consider the interaction $V_2$ that    contains 
 (among other terms),
\begin{eqnarray}
V_2 = -g^2 \sum_{c_i \sigma_i} T^{a}_{c_1 c_2} T^a_{c_4 c_3}  
 \int \left[ \Pi_i {d^3p_i\over{(2\pi)^3 2 p_i^+}} \right] 4
 \sqrt{x_1x_2x_3x_4} 
\delta_{\sigma_1 \sigma_2} \delta_{\sigma_4 \sigma_3 } \nonumber\\
{1\over{(x_1 -x_2)^2}} 
b_{p_1 c_1 \sigma_1}^{\dagger} d_{p_3 c_3 \sigma_3}^{\dagger}
b_{p_2 c_2 \sigma_2} d_{p_4 c_4 \sigma_4} + ...\label{Inst}
\end{eqnarray}
where $x_i \equiv p_i^+ /P^+$ is the fraction of the total $P^+$ carried
by the particle $i$.   Momentum conservation is assumed implicitly.
Similarly, $V_1$ is also  rewritten.

\subsection{Regularization}
As will be  illustrated in Section III, the ultraviolet (UV) and IR
divergences in the light-cone QCD are interwined (for additional
discussion see  Apendix E of ref. \cite{gluons}). For this reason 
and because at present only UV renormalization is achieved in a
systematic manner, details of regularization are important.  We
regulate the canonical quantum Hamiltonian with cutoffs on the changes
in transverse momenta and on longitudinal momentum fractions  as in ref.
\cite{gluons}, term by term in expansion of creation/annihilation
operators. For  example, an operator consisting of one creation and two
annihilation  operators would contain product of two regulators, one for
each of the annihilation  operators. Regulation of  operators containing
more than one creation operator (and their  Hermitian conjugates) is
slightly more involved. For details see \cite{gluons}. The instantaneous
 interaction
(\ref{Inst}) is viewed for regulation purposes as an exchange of a
 virtual particle between two vertices.

These general rules can be satisfied by
 various cutoffs. Specifically, we use a regulating factor of the form 
(arguments are generic for the moment)
 \begin{eqnarray}
r_{\Delta \delta} \left( k^{\perp \, 2},\  x \right) = \exp 
(-k^{\perp \,  2}/\Delta^2) r_{\delta}(x)  \label{R}  \   .
\end{eqnarray}
that is particularly convenient for analytical calculations. Here and in
 what follows
 $r_{\delta}$ is an infrared regulator that prevents its argument to
 become zero. We wish to
 study dependence on this regulator, hence we leave its functional form
 unspecified.
 $\Delta$ is an ultraviolet cutoff. The limit with the cutoffs removed
 is achieved by 
$\Delta \rightarrow \infty$, $\delta \rightarrow 0$, in which case 
$r_{\Delta \delta} \rightarrow 1$. In order to preserve kinematic 
symmetries,  we choose the arguments of the regulators to be  
kinematically invariant quantities, in particular, Jacobi momenta 
defined with respect to the {\it vertex} in question. As an example of a
regulated interaction, the standard quark-gluon  coupling has the 
following form in the regulated quantum theory,  
\begin{eqnarray}  \lefteqn{V_{1\, \Delta \delta} =   
\sum   \int  }\nonumber\\    & g& \left[ \, r_{\Delta \delta}\left( 
\kappa_{12}^{\perp \,2},\  {x_1\over{x_3}} \right)  r_{\Delta 
\delta}\left(\kappa_{12}^{\perp \,2},\  {x_2\over{x_3}}  \right)  
\,  \bar{u}_2 \left( \varepsilon_1^{\ast \mu }\gamma_{\mu} \right) u_3
 T^{c_1}_{c_2 c_3}   \,
 b^{\dagger}_2 a^{\dagger}_1 b_3      + {\rm h.c.} + ... \right] . 
 \label{v1reg}
 \end{eqnarray}
 For clarity, summation indices are implicit here, sum is over spins and
 colors of all 3
 particles,  integration runs over all 3-momenta with a momentum
 conserving $\delta$
  function, and  index $i$ is short for $(p_i (k_i), \sigma_i, c_i)$ as
 appropriate. $\bar{u}$, $u$ are
 light-cone spinors, $\varepsilon_{1}^{\mu}$ is a gluon polarization
 vector,
$\varepsilon_{k\sigma}^{\mu} = \left( \varepsilon^+ = 0,
 \varepsilon_{k\sigma}^-= 
2 k^{\perp} \cdot \varepsilon_{\sigma}^{\perp} /k^+,
 \varepsilon_{\sigma}^{\perp}\right)$, and
here $a_1^{\dagger}\equiv a_{k_1\sigma_1 c_1}^{\dagger}$ denotes
 specifically a gluon
 creation operator. 
(We did not intend to list all terms in $V_1$). The longitudinal 
fractions are 
${x_1 /{x_3}} \equiv {k_1^+ / {p_3^+}}$,  ${x_1/ {x_3}} 
\equiv {k_1^+ / {p_3^+}}$, 
satisfying
\begin{eqnarray}
x_1 + x_2 = x_3, 
\end{eqnarray} and the  transverse Jacobi  momentum 
$\kappa_{12}^{\perp}$ is defined by 
\begin{eqnarray}
k_1^{\perp}& = &{x_1\over{x_3}} p_3^{\perp} + \kappa_{12}^{\perp} 
\nonumber \\
p_2^{\perp}& = &{x_2\over{x_3}} p_3^{\perp} - \kappa_{12}^{\perp}\   \
.  \end{eqnarray}
With the specific choice of the functional form of the UV regulator in
(\ref{R}), the entire regulating factor in (\ref{v1reg})  is 
\begin{eqnarray}
\exp \left(-2\, {\kappa_{12}^{\perp \ 2}\over{\Delta^2}}\right)
r_{\delta}({x_1\over{x_3}} ) r_{\delta}({x_2\over{x_3}}) \    .
\end{eqnarray}

 This concludes our definition of the bare {\it regulated} Hamiltonian.

\subsection{Boost invariant similarity renormalization}

The dependence of the Hamiltonian on the ultraviolet (UV) regulator
 $\Delta$ is removed 
via similarity renormalization scheme whose basic idea has been outlined
 in the introductory part of this section. 
Similarity renormalization can be performed in various ways, in terms of
 matrix 
elements (the original G{\l}azek-Wilson formulation) or in terms of
 coefficients of creations and annihilations operators (G{\l}azek's
 similarity for particles), 
and for various definitions of the similarity formfactors in the
 effective Hamiltonian 
\cite{glazkovia,similarity}. For a review and comparison of numerical
 efficiency of various
 formulations for Hamiltonian in QCD in (2+1) dimensions see
 \cite{haridch}. The focus in this paper
 is restricted to the similarity renormalization
group for particles (i.e. for creation and annihilation operators)
that we use \cite{gluons}. 
For introduction to the similarity renormalization
group for particles  see the latest reference in
  \cite{glazkovia}.

Let us denote the regulated canonical bare light-front Hamiltonian, 
$H_{{\rm can}\, \Delta \delta}$, together with the as of yet
 undetermined counterterms, $X_{\Delta} $, in accordance with
 \cite{gluons} as 
$H_{\Delta \delta}$. 
$H_{\Delta \delta}$ is
 written in terms of creation/annihilation operators of bare, or
 current, quarks and gluons.
 For the purpose of this discussion, we denote {\it all}
 creation/annihilation operators, bosonic {\it and} fermionic,
 generically $a^{\dagger}, \  a$. 
 If this current Hamiltonian were to be 
used to describe mesons, physical states would have to be very
 complicated
 superpositions of current quarks and gluons. We wish to find an
 effective Hamiltonian 
that provides a simpler description at hadronic scales ${\lambda}$. To
 arrive to a simpler picture of mesons, the
 effective Hamiltonian should rather be expressed in terms of
 constituent quarks and gluons.
 Yet, the two Hamiltonians are equal,
 \begin{eqnarray}
 H_{\Delta \delta}(a)  = H_{\lambda}(a_{\lambda}), \label{equivH}
 \end{eqnarray}
 meaning that they have the same eigenvalues. ${\lambda}$ is a parameter
 of dimension 
 of mass that will be explained in more detail below. 
 As the Hamiltonians are related by the unitary
 transformation $U_{\lambda}$, so are the current and constituent
 particles,
 \begin{eqnarray}
 a_{\lambda} = U_{\lambda} \, a \, U_{\lambda}^{\dagger} \label{a}\  .
 \end{eqnarray}
 The creation/annihilation operators $a^{\dagger}$, $a$ satisfy the
 usual fermionic or bosonic commutation relations.  We use Lorentz
 invariant 
 normalization $2 (2\pi)^3 p^+ \delta^3(p'-p)$ for the commutation
 relations.
 
 Applying the transformation $U_{\lambda}$ to both sides of
 (\ref{equivH}) and owing to its unitarity, one obtains
 \begin{eqnarray}
U_{\lambda}^{\dagger} H_{\Delta \delta}\left(a\right) U_{\lambda} = 
H_{\lambda}\left(   U_{\lambda}^{\dagger} a_{\lambda}U_{\lambda}\right)
 \label{equivH2}
 \end{eqnarray}
 From (\ref{a}) follows that all dependence on $\lambda$ in
 (\ref{equivH2}) is  in the
 coefficients of the various combinations of creation/annihilation
 operators.  
 So differentiating with respect to $\lambda$ affects only the
 coefficients, not the operators, and we obtain
 \begin{eqnarray}
 {d\over{d \lambda}} {\cal H}_{\lambda} = 
 - \left[ \tau_{\lambda}\, , \, {\cal H}_{\lambda}\right]\label{simi1}
 \end{eqnarray}
where we have denoted 
 \begin{eqnarray}
 {\cal H}_{\lambda} \equiv H_{\lambda}\left(  U_{\lambda}^{\dagger}
 a_{\lambda}U_{\lambda}\right) = H_{\lambda}\left(a\right) 
 \end{eqnarray}
 in accordance with \cite{gluons}, and 
 $\tau_{\lambda} \equiv U_{\lambda}^{\dagger} U_{\lambda}'$ generates
 infinitesimal
 transformations. Equation (\ref{simi1}) is an operator equation, but
 apart from c-numbers 
 it contains only known, current, creation/annihilation operators and
 their commutators. For this reason, it can be solved in an operator
 form, in contrast to earlier formulations in terms of matrix elements.
 
 The transformation is constructed from the requirement that the
 Hamiltonian
$H_{\lambda}$ be band diagonal, which means, as we stated in the
 introductory part of this
section, that it does not contain direct couplings between arbitrary
 scales.
Parameter $\lambda$ is a measure of width of  momentum space formfactors
$f_{\lambda}$ that appear in vertices of the renormalized $H_{\lambda}$.
 For this reason,  $\lambda$ is  often referred to as {\it band width}. 
 Schematically, we require that 
 \begin{eqnarray}
 {\cal H}_{\lambda} = f_{\lambda} G_{\lambda} \label{hg}
 \end{eqnarray}
 which indicates that we want ${\cal H}_{\lambda}$ to vanish in the part
 of the phase space were the {\it similarity formfactor} $f_{\lambda}$
 is zero. This equation also defines $G_{\lambda}$\footnote{Be aware
 that G{\l}azek in \cite{gluons} uses notation 
 ${\cal G}_{\lambda}$ to make a distinction with his earlier works.}.
 
  It is easier to work with $G_{\lambda}$ than with 
 ${\cal H}_{\lambda}$\footnote{Factorization of the overall
 $f_{\lambda}$, 
 and writing the renormalization group equations for $G_{\lambda}$
 rather than
 $H_{\lambda} $ is not necessary, but it sure makes the equations much
 easier to work with.
  For comparison, see the earliest reference in \cite{similarity}.}.
 $G_{\lambda}$ is split into the free part  $G_{0\lambda}=G_0$ that
 is, in present work, independent of interactions and does not change
 with $\lambda$, 
 and the interacting 
 part $G_{I\lambda}$. Rewriting the differential eqn. (\ref{simi1})
 gives
 \begin{eqnarray}
 {df_{\lambda}\over{d \lambda}} {G}_{I\lambda} + f_{\lambda} {d\over{d
 \lambda}} 
 {G }_{I\lambda}= 
 -\left[ \tau_{\lambda}\, , \, G_{0}\right]
 - \left[ \tau_{\lambda}\, , \, G_{I\lambda}\right] \label{simi2}
 \end{eqnarray}
  The left side of the equation is independent of $G_{0}$ because 
 $G_{0}$
does not change with $\lambda$ and because $f_{\lambda}$ does not 
change on the diagonal.  
  $G_{I\lambda}$ consists of the canonical interactions as well as new
 effective interactions
ensuring that the effective Hamiltonian is equivalent to the original
 current one. Therefore,
 (\ref{simi2}) describes two unknowns, $\tau_{\lambda}$ and
 $G_{I\lambda}$. 
 We have  freedom to arbitrarily split this equation into two. Without
 loss of generality, we
can assume that when the interactions vanish, so does the generator
 $\tau_{\lambda}$.
 If we choose
 \begin{eqnarray}
   f_{\lambda} {d\over{d \lambda}} 
 {G }_{I\lambda}= -  f_{\lambda}\left[ \tau_{\lambda}\, , \,
 G_{I\lambda}\right] ,\label{simi3}
\end{eqnarray}
 then 
 \begin{eqnarray}
  \left[ \tau_{\lambda}\, , \, G_{0}\right]
   ={d\over{d \lambda}} \Big( (1-f_{\lambda})G_{I\lambda}\Big) ,
 \label{simi4}
\end{eqnarray} 
 and the generator of the transformation is of order $G_{I\lambda}$.
 Changes in 
 $G_{I\lambda}$, corresponding to  new effective terms, are of second
 order in interactions.
 
 Equation (\ref{simi4}) can be solved for $\tau_{\lambda}$, for example
 order by order 
 in the interactions. In fact, with $G_0$ being the free Hamiltonian
 (\ref{h0}), the solution is
simple. It is the same operator as the right hand side of the equation
 (\ref{simi4}) with an
additional factor $\left[ \sum_{a} E_i - \sum_{c}E_j\right]^{-1}$, where
 $a$, $c$ stands for
annihilated/created, and $E$'s are free light-front energies
 created/annihilated in the vertex
(see (\ref{h0})). The additional factor arises from the free energy in
 $G_0$ and 
contracting one creation/annihilation operator with the conjugate in
 $G_0$.  
It is, however, not necessary to explicitly find $\tau_{\lambda}$.
If we {\it denote} the solution to (\ref{simi4}) 
 \begin{eqnarray}
 \tau_{\lambda} = \left\{ {d\over{d \lambda}} \left(
 (1-f_{\lambda})G_{I\lambda}\right)
 \right\}_{G_{0}}
 \end{eqnarray}
 then the solution to (\ref{simi3}) subject to the boundary condition
 \begin{eqnarray}
  {G }_{\lambda \rightarrow \infty }= H_{\Delta \delta},
 \end{eqnarray} 
 where $H_{\Delta \delta}$ includes both known canonical terms and
 unknown  counterterms, can be written as
 \begin{eqnarray}
 {G }_{\lambda}= H_{{\rm can} \, \Delta \delta} 
 + X_{\Delta}+\int_{\infty}^{\lambda} d\sigma 
 \left[  f_{\sigma} G_{I\sigma}\, , \, \left\{ {d\over{d \sigma}} \Big(
  (1-f_{\sigma})G_{I\sigma}\Big) \right\}_{G_{0}}\right] \label{glambda}
   \end{eqnarray}
Counterterms $X_{\Delta}$ should remove any dependence on $\Delta$
 arising from the
 second term. This criteria is sufficient to determine $X_{\Delta}$.
 Finite parts of the
second term constitute new effective interactions and/or modifications
 to canonical terms.
Finally, the Hamiltonian is found by substitution to (\ref{hg}).

The equation  (\ref{glambda}) is an operator equation but the
 creation/annihilation operators
 are not affected by differentiation. They merely play a role for
 "counting" purposes. If we are
  interested in writing down a differential equation for a coefficient
 of a specific
   combination of creation/annihilation operators, to any given order in
 $g$ we can see what
 operators in the interaction Hamiltonian need to be included on the
 right-side of the
 equation, and which of the creation/annihilation operators on the right
 side need to be
 contracted. Note that since the right side of (\ref{glambda}) is
 proportional to a
 commutator, the coefficients in the effective Hamiltonian arise only
 from connected
 terms.

The preceding discussion was valid for an arbitrary choice of the
 similarity formfactor, and we have not specified  how the formfactors
 are imposed. 
Specifically, we require that any combination of creation/annihilation
 operators in the
Hamiltonian is accompanied by a formfactor $f_{\lambda}$;
a particularly useful choice of the similarity formfactor 
 $f_{\lambda}$  has the following form 
 \begin{eqnarray}
 f_{\lambda}({\cal M}_{c}^2 -{\cal M}_{a}^2) \equiv f_{\lambda}(ca) =
\exp \left( -  {[{\cal M}_{c}^2 -{\cal M}_{a}^2]^2\over{\lambda}^4}
\right) \label{fl}
\end{eqnarray}
where ${\cal M}_{c}^2 \equiv\left( \sum_{i \  created} p_i\right)^2$ is
the  square of sum of
all free four-momenta {\it created} in the vertex and similarly,
 ${\cal M}_{a}^2
\equiv\left( \sum_{j \  annih. } p'_j\right)^2$ is the square of all
 free four-momenta 
{\it annihilated} in the vertex. 

 For example, for any finite  $\lambda$ Hamiltonian, the integrand of
 the operator (\ref{v1reg})
  is modified by insertion of a factor 
 \begin{eqnarray}
 f_{\lambda}( (k_1+p_2)^2 -p_3^2) = \exp \left( - {[(k_1+p_2)^2 -
 m^2]^2\over{\lambda}^4}
  \right) .
 \end{eqnarray}
The corresponding term in 
$G_{\lambda}$, i.e. the Hamiltonian without the vertex formfactor, is to
 the lowest order in interactions just equal to (\ref{v1reg}) and 
 independent of $\lambda$. 

In this work, we expand the Hamiltonian relevant for a quark-antiquark
 state in powers of $g$,  and find effective interactions up to $g^2$. 
The perturbation theory for
 $G_{\lambda}$ is well explained in section 3.c of reference
 \cite{gluons}. As it is often the case, in practice the most convenient
 way to determine the factors arising from the similarity transformation
 from infinity down to the finite scale $\lambda$ in  eqn.
 (\ref{glambda}) is   with the help of  diagrammatic rules.
 Unfortunately, these seem to be the best kept secret in the field, and
 cannot be found listed in any reference, perhaps for fear of confusion
 with the ordinary perturbation theory for the S-matrix
 \cite{glazkovia}.  The calculation of effective interactions to ${\cal
 O}(g^2)$  is much simpler than at higher orders, because it involves
 only  $G_{\lambda}$ at ${\cal O}(g)$ which is independent of the
 similarity scale. Consequently, all dependence on the scale in eqn.
 (\ref{glambda}) is in $f_{\sigma}$ and $d \, f_{\sigma}/d\sigma$. This
 is not true in general, when care must be taken to keep track of all
 $\sigma$ dependence. Therefore, the second order calculation does not
 require  ``full-blown''  diagrammatic rules and we will not develop
 them here.  We will, however, show the simplified rules sufficient for
 the present calculation, with the understanding that they should and
 can  be generalized for higher order applications.

To draw a diagram for the effective coefficient of any operator ${\cal
 O}_{\lambda}$  in $G_{\lambda}$,
\begin{itemize}
\item draw an external line for each creation and each annihilation
 operator in ${\cal O}_{\lambda}$. Draw all {\it connected} diagrams
 connecting the external lines. (At second order, one-body operator has
 only one diagram, Fig. 1, reminiscent of a self-energy diagram in the
 ordinary perturbation theory. The two-body operator has two.) 
\item Let us denote total momentum annihilated/created in a vertex  $i$
 as $a_i, \ c_i$. (At second order, $i=1,2$.) Let $p_i^+, p_i ^{\perp}$
 be the  momentum {\it conserved in the vertex} $i$, and light-cone free
 energies $E_{c_i} \equiv ({\cal M}^2_{c_i}+p_i^{\perp\, 2})/p_i^+,$
 created, and $E_{a_i} \equiv ({\cal M}^2_{a_i}+p_i^{\perp\, 2})/p_i^+,$
 annihilated.

At each vertex $i$, include the standard factors, as well as regulators,
  already present in the $G_{\sigma}$, such as,  for example, the
 familiar
$\bar{u}_i \epsilon^{\mu}\gamma_{\mu} u$,with a
 momentum conserving delta function, at each $qqg$ vertex $i$.  

\item The novel element comes from  the similarity formfactors and
 scale.  The entire diagram is to be integrated 
 $\int_{\infty}^{\lambda}d\, \sigma$. At second order, 
assign 
\begin{eqnarray}
\left[\left( {d\over{d\,\sigma}} f_{\sigma}(c_1 a_1)\right)
 {1\over{E_{c_1} - E_{a_1}}} 
f_{\sigma}(c_2 a_2)-
\left( {d\over{d\,\sigma}} f_{\sigma}(c_2 a_2)\right) {1\over{E_{c_2} -
 E_{a_2}}} 
f_{\sigma}(c_1 a_1)\right]
\end{eqnarray}
to the diagram. This factor contains all the dependence on the scale,
 and, for $f_{\sigma}$ given in (\ref{fl}), it can simply be integrated,
 resulting in 
\begin{eqnarray}
\left[ {p_1^+ ({\cal M}_{c_1}^2 - {\cal M}_{a_1}^2) + p_2^+ ({\cal
 M}_{c_2}^2 - {\cal M}_{a_2}^2) \over{ ({\cal M}_{c_1}^2 - {\cal
 M}_{a_1}^2)^2 + ({\cal M}_{c_2}^2 - {\cal M}_{a_2}^2)^2}}\right] \left[
 f_{\lambda}(c_1 a_1)f_{\lambda}(c_2 a_2) -1 \right]. \label{fsim2}
\end{eqnarray}
\item Diagrams can acquire an overall sign and/or combinatorial factors
 so that the same ordering of uncontracted creation/annihilation
 operators on both sides of the eqn. (\ref{glambda}) is achieved
 \cite{glazkovia}. At second order, however, these overall factors are
 obvious. 
\item Generalization of the procedure so that it is applicable for
 higher order calculations is straightforward.
\end{itemize}

As a specific example, we present in the next section the
 renormalization of the one body operator.
 
\begin{figure}
\begin{center}
\psfig{figure=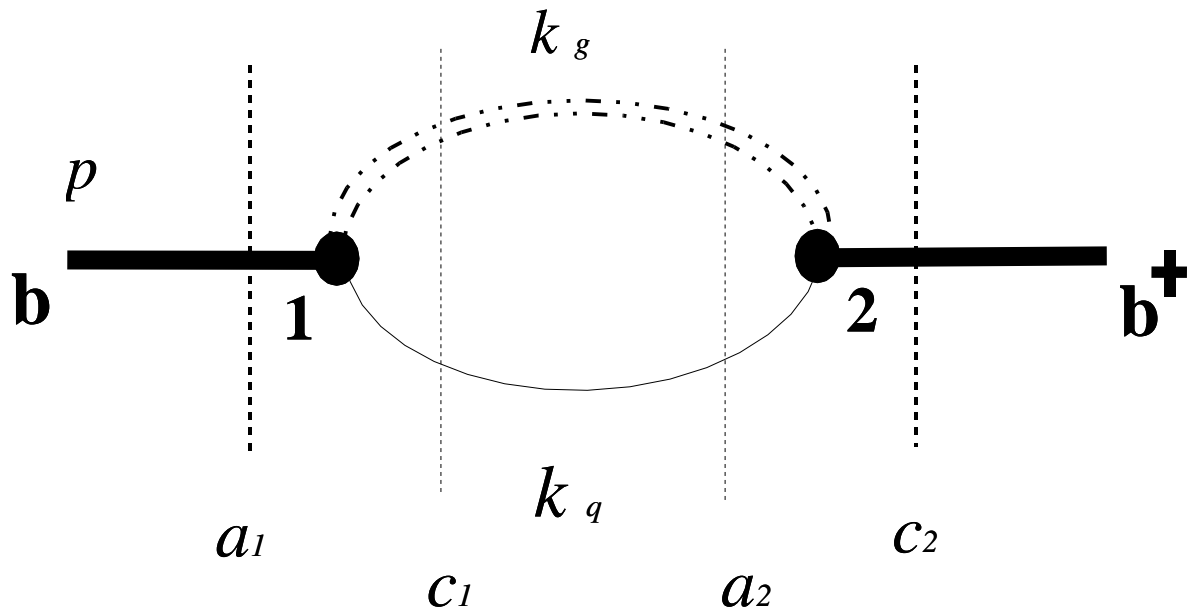}
\end{center}
\caption{Diagram for the quark one body operator at second order in $g$,
 not to be confused with the ordinary perturbation theory for $S$
 matrix. Thick lines correspond to uncontracted creation/annihilation
 operators. Double line is a contracted gluon, single thin line is a
 contracted quark. Vertical dashed lines  show which momenta add up to
 $a_i$, $c_i$, respectively. All other notation is explained in the
 text.} 
\end{figure}

\section{Renormalization of the one-body operator}

The one-body operator was previously derived, but not further studied,
in ref. \cite{glazkovia} with slightly different choices. 

The one-body operator when $\lambda=\infty$ is
\begin{eqnarray}
\int \left[ {\rm d}^3 p \right] \left( {p^{\perp  2}+m_{\infty}^2 
\over{p^+}} +{\delta m_{\Delta}^2\over{p^+}} \right) \left( b^{\dagger}
 b 
+d^{\dagger} d\right)
\label{onebody}
\end{eqnarray}
where $m_{\infty}^2$ is the bare mass, $\Delta$ is the UV regulator  and
$\delta m_{\Delta}^2$ is the counterterm to be 
determined so that renormalized Hamiltonian is independent of the
 regulator.

When $\lambda$ is finite, there are corrections 
(with coefficient $\delta m_{\lambda}^2/p^+$)
to the one-body operator\footnote{With boost invariant similarity, 
corrections  have  the proper dispersion relation.} 
which start at ${\cal O}\left(g^2\right)$. 

To visualize the diagrammatic approach outlined at the end of previous
 section, we present in Fig. 1 a diagram for the lowest order correction
 to the one body operator with the notation used in the list of rules.
 For example, momentum $a_1$ is the free momentum of the uncontracted
 quark $p$; momentum $c_1$ in this diagram is equal to sum of the free
 momentum of the virtual gluon $k_g$ and the virtual quark $k_q$. Three
 momenta are conserved, so only one of the virtual particles momenta is
 independent.

One factor of $(\ref{v1reg})$ (with appropriate momenta and without
 integration over the external momentum) stands for  each vertex. Sums
 of the polarisation vector, spin and colors of the virtual particles
 lead to standard expressions involving color factors and matrix
 elements of 
$ D_{\mu \nu} (k_g) \gamma^{\mu} \left( / \! \! \! k_q + m \right)
 \gamma^{\nu}$  in the spinors of the uncontracted quark. It is
 convenient to express the virtual momenta in terms of Jacobi momenta
 $\kappa^{\perp }, \  x$. This loop momentum is integrated over. 

The  similarity factor (\ref{fsim2}) further simplifies. Using that
 $p_1^+ = p_2^+= p^+$, and 
${\cal M}_{c_1}^2= {\cal M}_{a_2}^2=m^2,$  \  
${\cal M}_{c_2}^2= {\cal M}_{a_1}^2=(m^2+ \kappa^{\perp \,2})/x +
 \kappa^{\perp \,2}/(1-x)$ in this case, we get
\begin{eqnarray}
{p^+\over{m ^2 -({m^2+ \kappa^{\perp \,2}\over{x}} + {\kappa^{\perp
 \,2}\over{1-x}})}}
\left[ f_{\lambda}^2\left(m ^2 -({m^2+ \kappa^{\perp \,2}\over{x}} +
 {\kappa^{\perp \,2}\over{1-x}}) \right) -1\right].
\end{eqnarray}
In order to write the one body operator with the same ordering as the
 original mass term, the uncontracted $b, \ b^{\dagger}$  have to be
 commuted  which gives an overall (-1). 

After this straightforward algebra, we find that the correction to the
 one body operator is indeed of the expected form   $constant/p^+ \
 b^{\dagger} b$ (independent of $p^{\perp}$) 
with the coefficient
\begin{eqnarray}
{g^2\over{(4\pi)^2}}{N^2-1\over{2N}}\int_0^1 {\rm d}y \ 
r_{\delta}^2(y) \  r_{\delta}^2(1-y)\int_0^{\infty}{\rm d}z\left[
2y + 4 {1-y\over{y}} - {4m^2\over{z+
 {y\over{1-y}}m^2}}\right]\nonumber\\
\exp\left(-2 {z y(1-y)\over{\Delta^2}}\right) 
\left[ \exp \left(-{2\over{\lambda^4}} (z+
 {y\over{(1-y)}}m^2)^2\right)-1\right] \label{dml}
\end{eqnarray}
independent of $p^+$.
Here $m\equiv m_{\infty}$; $z$ arises from integration over transverse
 momenta and carries a
 dimension of mass squared. 
The coefficient (\ref{dml}) consists of  terms
 $ \mu^2_{\Delta}$ independent of
 $\lambda$, containing UV/IR divergences, and $\lambda$-dependent terms,
 both finite 
 $\mu^2_{\lambda}$
  and IR divergent $\mu^2_{\lambda, \ \delta}$, 
  \begin{eqnarray}
  {g^2\over{(4\pi)^2}}{N^2-1\over{2N}}\left[\mu^2_{\Delta} 
  +\mu^2_{\lambda}+\mu^2_{\lambda , \ \delta}\right],
 \end{eqnarray} 
 \noindent  where:
\begin{eqnarray}
\mu^2_{\Delta} & = &
-\Delta^2 \  c_{\delta} + 4 m^2\log{\Delta^2\over{m^2}}
 -4m^2 (\gamma + \log 2 -2) \\
 \mu^2_{\lambda} &= &\lambda^2 \int_0^1 {\rm d}y \int_0^{\infty} 
 {{\rm d}z\over{\lambda^2}}
\exp \left(-{2\over{\lambda^4}} (z+ {y\over{(1-y)}}m^2)^2\right)
\left\{2y - 
{4m^2\over{z+ {y\over{1-y}}m^2}}\right\} \nonumber\\
& -& 
\lambda^2 \sqrt{2\pi}\left[ 1+ \int_0^1 {\rm d}y {1-y\over{y}}
{\rm Erf }\left( \sqrt{2} {m^2\over{\lambda^2}}{y\over{1-y}}\right)
 \right]\\
\mu^2_{\lambda, \ \delta}& = & \lambda^2 \sqrt{2\pi} \int_0^1 
{\rm d}y {r_{\delta}^2(y)\over{y}} . \label{onebodydeltadiv}
\end{eqnarray}
Note that the coefficient of the quadratic divergence 
$c_{\delta} \equiv \int_0^1{\rm d}y[r_{\delta}^2(1-y) 
/(1-y) + 2 r_{\delta}^2(y)/y^2]$ diverges as $\delta \rightarrow 0$. 

The one-body operator can be written as
\begin{eqnarray}
\int \left[ {\rm d}^3 p \right] \left( {p^{\perp  2}+m_{\lambda}^2 
\over{p^+}}  \right) \left( b^{\dagger} b +d^{\dagger} d \right)
\label{onebodylambda}
\end{eqnarray}
where
\begin{eqnarray}
m_{\lambda}^2 & = & m_{\infty}^2 + \delta m_{\Delta}^2
 +{g^2\over{(4\pi)^2}}{N^2-1\over{2N}}\left[\mu^2_{\Delta} 
  +\mu^2_{\lambda}+\mu^2_{\lambda , \ \delta}\right] \nonumber\\
  & \equiv& m^2(\lambda)  +
 {g^2\over{(4\pi)^2}}{N^2-1\over{2N}}\mu^2_{\lambda , \ \delta}
\label{m}
\end{eqnarray}
The UV divergent $\mu^2_{\Delta}$ is combined with the counterterm
$ \delta m_{\Delta}^2$ so that the dependence on the regulator $\Delta$
 is removed. 
Nevertheless, the mass $m_{\lambda}^2$ is not finite in the limit
 $\delta \rightarrow 0$.
In what follows, we are concerned only with  the behavior in this limit,
rather than details of the finite terms. 
For this reason and for simplicity, we grouped all finite terms
 contributing to
 $m_{\lambda}^2$ into $m^2(\lambda)$. 

\section{Two-body operators}
We are seeking a coefficient $\cal V$ of an operator consisting of two 
creation and two annihilation operators, 
\begin{eqnarray}
\int \, 4 \sqrt{x_1 x_2 x_3 
x_4} \  {\cal V}_{\lambda} \ f_{\lambda}({\cal M}_{12}^2 -{\cal
 M}_{34}^2) \, 
b^{\dagger}_3  d^{\dagger}_4 b_1 d_2
\end{eqnarray}
that creates/annihilates quark-antiquark pair with free momenta 
\begin{eqnarray}
{\rm created:} \   %&  & \nonumber\\
p_3& =&  \left( (m^2 +\kappa_{34}^2)/x_3, \ x_3, \  \kappa_{34}^{\perp} 
\right) \\
p_4& =&  \left( (m^2+ \kappa_{34}^2)/x_4, \ x_4=1-x_3, \  
-\kappa_{34}^{\perp} \right)
\\
{\rm annihilated:}\   % &  & \nonumber\\
p_1& =&  \left( (m^2+ \kappa_{12}^2)/x_1, \ x_1, \  \kappa_{12}^{\perp} 
\right) \\
p_2& =&  \left( (m^2+\kappa_{12}^2)/x_2, \ x_2=1-x_1, \ 
 -\kappa_{12}^{\perp} 
\right)
\end{eqnarray}
which we expressed in terms of relative (Jacobi) momenta and set the
 total 
transverse momentum of the initial and final two-body states to zero
 because 
of kinematical boost invariance. We also factored out the total $P^+$
 because 
all regulators and form factors are expressed in terms of boost
 invariant 
quantities. The conservation of the three momentum is implicit here and
 in what 
follows. 

The coefficient ${\cal V}_{\lambda}$ depends on momenta, and contains
 terms of 
${\cal O}(g^2)$ and higher.\footnote{We factored out the omnipresent $4 
\sqrt{x_1 x_2 x_3 x_4}$ because it cancels with the similar factor in
 the 
definition of the bound state wavefunction.}
We are interested only in the color singlet and most IR divergent
 part of this operator. (This part is diagonal in spins).
  Depending on its degree of divergence, it can be concluded
 whether or not the subleading terms are IR convergent. 

The canonical Hamiltonian already contains a two-body operator of this
 form, the 
instantaneous gluon exchange $V_2$ (\ref{Inst}). With the infrared
 regularization as in 
\cite{gluons} its coefficient is
\begin{eqnarray}
\lefteqn{- g^2 C_F{\cal V}_{\rm inst} \equiv - g^2 C_F
 {1\over{(x_1-x_3)^2}} 
%4 \sqrt{x_1 x_2 x_3 x_4}
}\nonumber\\
& &\left\{
\theta (x_1-x_3)\, 
r_{\delta}\left({x_3\over{x_1}}\right)\, 
r_{\delta}\left({x_1-x_3\over{x_1}}\right)\,
r_{\delta}\left({x_2\over{x_4}}\right)\, 
r_{\delta}\left({x_1-x_3\over{x_4}}\right)
 \right. \nonumber\\
& &\left.+ \  \theta (x_3-x_1)
r_{\delta}\left({x_1\over{x_3}}\right)\, 
r_{\delta}\left({x_3-x_1\over{x_3}}\right)\,
r_{\delta}\left({x_4\over{x_2}}\right)\, 
r_{\delta}\left({x_3-x_1\over{x_2}}\right)
\right\} \label{vinst}
\end{eqnarray}
where $r_{\delta}$ is the (unspecified) infrared regulator.

With the choice of $f_{\lambda}$ (\ref{fl}) and regularization as in
 ref. 
\cite{gluons}, the spin independent, most infrared divergent, color
 singlet part 
of the 
coefficient of the effective two-body operator to lowest (i.e. second)
 order in 
coupling is
\begin{eqnarray}
 -g^2 C_F\,\left( {\cal V}_{\rm 1}+{\cal V}_{\rm 2}\right)
 \end{eqnarray}
 where
\begin{eqnarray}
\lefteqn{-g^2 C_F\, {\cal V}_{\rm 1} \equiv -g^2 C_F 
%4 \sqrt{x_1 x_2 x_3 x_4}
{k_5^{\perp 2}\over{(x_1-x_3)^2}}\left\{
{\theta (x_1-x_3)\over{(x_1-x_3)}} \right. }\nonumber\\
& &\left[ f_{\lambda}({\cal M}_{35}^2 -{\cal M}_1^2)\, f_{\lambda}({\cal
M}_{25}^2 -{\cal M}_4^2)-1\right]
{x_4 ({\cal M}_{25}^2 -{\cal M}_4^2) +x_1 ({\cal M}_{35}^2 -{\cal
 M}_1^2)
\over{({\cal M}_{25}^2 -{\cal M}_4^2)^2+({\cal M}_{35}^2 -{\cal
 M}_1^2)^2}}
\nonumber\\
& & \left.
r_{\delta}\left({x_3\over{x_1}}\right)\, 
r_{\delta}\left({x_1-x_3\over{x_1}}\right)\,
r_{\delta}\left({x_2\over{x_4}}\right)\, 
r_{\delta}\left({x_1-x_3\over{x_4}}\right)
\right\}   \  , \label{v1}
\\
\nonumber\\
%\left.+\ 
\lefteqn{-g^2 C_F \, {\cal V}_{\rm 2} \equiv -g^2 C_F 
%4 \sqrt{x_1 x_2 x_3 x_4}
{k_5^{\perp 2}\over{(x_1-x_3)^2}}\left\{ {\theta 
(x_3-x_1)\over{(x_3-x_1)}}\right.} \nonumber\\
&  &\left[ f_{\lambda}({\cal M}_{15}^2 -{\cal M}_3^2)\,
 f_{\lambda}({\cal 
M}_{45}^2 -{\cal M}_2^2)-1\right]
{x_2 ({\cal M}_{45}^2 -{\cal M}_2^2) +x_3 ({\cal M}_{15}^2 -{\cal
 M}_3^2)
\over{({\cal M}_{45}^2 -{\cal M}_2^2)^2+({\cal M}_{15}^2 -{\cal
 M}_3^2)^2}}
\nonumber\\
&  &\left.
r_{\delta}\left({x_1\over{x_3}}\right)\, 
r_{\delta}\left({x_3-x_1\over{x_3}}\right)\,
r_{\delta}\left({x_4\over{x_2}}\right)\, 
r_{\delta}\left({x_3-x_1\over{x_2}}\right)
\right\} \label{v2}
\end{eqnarray}
It arises from contracting one pair of creation and annihilation
 operators 
of a gluon with  
\begin{eqnarray}
k_5 = \left( k_5^{\perp \ 2}/x_5, x_5, k_5^{\perp} \equiv 
\kappa_{34}^{\perp} -\kappa_{12}^{\perp}\right)
\end{eqnarray}
\noindent where $x_5 =|x_3-x_1|$ corresponding to the two
 time-orderings. 
Invariant masses $\cal M$ are defined as 
\begin{eqnarray}
{\cal M}_{ij}^2 \equiv (p_i +p_j)^2 \\
{\cal M}_{i}^2 \equiv (p_i)^2 =m^2
\end{eqnarray}
at each vertex 
and $r_{\delta}$ are the infrared regulators, in accordance with 
\cite{gluons}. ${\cal V}_{\rm 1}$ corresponds to particle 1 emitting the
 gluon 
that is absorbed by particle 2; 
in ${\cal V}_{\rm 2}$ particle 2 emits gluon, particle 1 absorbs it.
The second line in both expressions comes from integrating the width
 parameter 
in the similarity RG procedure from infinity down to its value
 $\lambda$.

In terms of the Jacobi momenta, the arguments of the formfactors are
\begin{eqnarray}
{\cal M}_{25}^2 -{\cal M}_4^2 & = & x_2 {k_5^{\perp 2}\over{x_5}} - 2 
\kappa_{12}^{\perp} \cdot k_5^{\perp} + {x_5 \over{x_2}}(m^2
 +\kappa_{12}^{\perp 
 2}) \\
{\cal M}_{35}^2 -{\cal M}_1^2 & = & {x_1^2\over{x_3}} {k_5^{\perp
 2}\over{x_5}} 
+ 2  {x_1\over{x_3}}\kappa_{12}^{\perp} \cdot k_5^{\perp} + {x_5
 \over{x_3}}(m^2 
+\kappa_{12}^{\perp  2}) \\
{\cal M}_{15}^2 -{\cal M}_3^2 & = & x_1 {k_5^{\perp 2}\over{x_5}} - 2 
\kappa_{12}^{\perp} \cdot k_5^{\perp} + {x_5 \over{x_1}}(m^2
 +\kappa_{12}^{\perp 
 2})  \\
{\cal M}_{45}^2 -{\cal M}_2^2 & = & {x_2^2\over{x_4}} {k_5^{\perp
 2}\over{x_5}} 
+ 2  {x_2\over{x_4}}\kappa_{12}^{\perp} \cdot k_5^{\perp} + {x_5
 \over{x_4}}(m^2 
+\kappa_{12}^{\perp  2}) \label{arg}
\end{eqnarray}

\section{Infrared divergences}
The coefficients of the effective one-body and two-body operators
 diverge when the infrared
 regulator is removed, i.e. $r_{\delta} \rightarrow 1.$  In previous
 calculations \cite{P2}
the infrared divergences were found to cancel in the color singlet $q
 \bar{q}$ states, leaving
 however, the nonsinglet states with infinite mass. The remaining part
 of the two-body
 potential was found to be confining. A similar observation was made
 recently 
 regarding  QCD in (2+1) dimensions \cite{haridch}. We want to check
 whether the
 cancelation occurs here.

It is easier to address the infrared divergence issue
in the framework of a bound state equation, or equivalently, in terms of
 expectation values of
 the effective Hamiltonian. In that case, the creation and 
annihilation operators are contracted, and the integrands of the
 momentum 
integrals are just c-numbers consisting of the coefficients of the
 operators and 
wavefunctions of the bound state. 
The bound state eqn. reads
\begin{eqnarray}
\lefteqn{\left( 4m^2 + 4m \, E\right) \Phi_{12}=}\nonumber\\
& & \left[\kappa_{12}^2 +m^2(\lambda)+{g^2\over{(4 
\pi)^2}}\, 
{N^2-1\over{N}}\, \mu^2_{\lambda , \  \delta}  
\right]\left({1\over{x_1}} + 
{1\over{x_2}}\right) \Phi_{12} \nonumber\\
& & - {g^2 \over{4 \pi^2}} C_F 
\int {{\rm d}x_3 {\rm d}^2\kappa_{34}^{\perp} \over{\pi}}
\left[ {\cal V}_1 + {\cal V}_2 + {\cal V}_{\rm inst}
 \right]f_{\lambda}({\cal 
M}_{12}^2 -{\cal M}_{34}^2) \Phi_{34}
\label{boundsteqn}
\end{eqnarray}
where $ \left( 4m^2 + 4m \, E\right)$ is the 
eigenvalue\footnote{We use this definition of eigenvalue for its
 convenience in 
nonrelativistic limit, when $4m^2$ cancels with the same expression on
 the other 
side of the equation, and after dividing by $4m$ the bound state
 equation 
reduces to the usual Schrodinger equation for eigenvalue $E$.}, 
and $\Phi_{ij} = \Phi(x_i, \kappa_{ij}^{\perp})$ is the bound state wave
function of a  quark-antiquark  state of total momentum 
$P \equiv \left(P^+, \, P^{\perp}=0 \right)$: 
\begin{eqnarray}
\vert P \rangle = \int {{\rm d}x_i {\rm d}^2\kappa_{ij}^{\perp}
 \over{2(2 \pi)^3 x_i x_j}}
\sqrt{x_i x_j}\, \Phi(x_i, \kappa_{ij}^{\perp}) \,
 b_i^{\dagger}d_j^{\dagger} \vert 0 \rangle 
\  \  \   .
\end{eqnarray}
The first line of the right side of the bound state eqn.
 (\ref{boundsteqn})
 comes from one-body operators, the 
integral part arises from the two-body operator. 

The infrared structure of the expectation value of the two-body operator
 depends
on the similarity formfactors in the various coefficients $\cal V$. The
 coefficients also
 contain two different sets of infrared regulators corresponding to the
 two time orderings.
To make the dependence on all formfactors explicit, let us introduce
 $v_1, \ v_2 $  as
\begin{eqnarray}
{\cal V}_{\rm 1} \equiv \left[ f_{\lambda}({\cal M}_{35}^2 -{\cal
 M}_1^2)\, 
f_{\lambda}({\cal 
M}_{25}^2 -{\cal M}_4^2)-1\right] \theta (x_1-x_3) \, v_{\rm 1}
 \label{malev1} \\
{\cal V}_{\rm 2} \equiv \left[ f_{\lambda}({\cal M}_{15}^2 -{\cal
 M}_3^2)\, 
f_{\lambda}({\cal 
M}_{45}^2 -{\cal M}_2^2)-1\right] \theta (x_3-x_1)  \, v_{\rm 2}
 \label{malev2}
\end{eqnarray}
${\cal V}_{\rm inst}$ does not contain any formfactor, but it does
 contain the two 
sets of infrared regulators for $x_3<x_1$ and $x_3>x_1$. It is
 convienient to 
introduce $v_{\rm inst \, 1}$, $v_{\rm inst \, 2}$ in analogy with $v_1,
 \ v_2 \,$: 
\begin{eqnarray}
{\cal V}_{\rm inst}\equiv \theta (x_1-x_3) \,  v_{\rm inst \, 1} +\theta
(x_3-x_1)  \, v_{\rm inst \, 2}
\end{eqnarray}
This defines $v$'s. Note that the infrared regulators in $v_{\rm 1}$ and
$v_{\rm inst \, 1}$ are the same; so are in $v_{\rm 2}$ and
$v_{\rm inst \, 2}$.

To extract the infrared structure of the integral in 
(\ref{boundsteqn}), we first note that the IR divergence in the
 coefficients 
(\ref{vinst}) and  (\ref{v1}),
 (\ref{v2}) occurs when $x_5 \rightarrow 0$ and that the small $x_5$
  behavior of $v_{\rm inst \, i}$ has opposite sign compared to $v_{\rm
 i}$.
For this reason we first regroup the integrand in 
(\ref{boundsteqn}) by adding and subtracting
$v_{\rm inst \, i}$ multiplied by the same formfactors $f_{\lambda}$ as
 those multiplying
 $v_{\rm i}$ in (\ref{malev1}), (\ref{malev2}). This splits the integral
 into two parts, 
\begin{eqnarray}
- {g^2 \over{4 \pi^3}} C_F \int {\rm d}x_3 {\rm d}^2\kappa_{34}^{\perp} 
\left[ {\cal V}_1 + {\cal V}_2 + {\cal V}_{\rm inst}
 \right]f_{\lambda}({\cal 
M}_{12}^2 -{\cal M}_{34}^2) \Phi_{34} = I + {\rm I} \  ,
\end{eqnarray}
where
\begin{eqnarray}
\lefteqn{ I = - {g^2 \over{4 \pi^3}} C_F \int {\rm d}x_3 {\rm 
d}^2\kappa_{34}^{\perp}}\nonumber\\ 
& & \left\{ \theta (x_1-x_3)
\left( v_{\rm 1} - v_{\rm inst \, 1}\right)\left[ f_{\lambda}({\cal
 M}_{35}^2 
-{\cal M}_1^2)\, f_{\lambda}({\cal 
M}_{25}^2 -{\cal M}_4^2)-1\right] 
\right. \nonumber\\
& & \left. + \theta (x_3-x_1)  \, \left( v_{\rm 2}- v_{\rm inst \,
 2}\right)
\left[ f_{\lambda}({\cal M}_{15}^2 -{\cal M}_3^2)\, f_{\lambda}({\cal 
M}_{45}^2 -{\cal M}_2^2)-1\right]\right\}\nonumber\\
&  & f_{\lambda}({\cal M}_{12}^2 -{\cal M}_{34}^2) \Phi_{34} \  ,
\label{conv}\end{eqnarray}
and 
\begin{eqnarray}
{\rm I} = - {g^2 \over{4 \pi^3}} C_F \int {\rm d}x_5 {\rm
 d}^2k_5^{\perp} 
\left\{ \theta (x_1-x_3) \, v_{\rm inst \, 1} f_{\lambda}({\cal
 M}_{35}^2 -{\cal 
M}_1^2)\, f_{\lambda}({\cal M}_{25}^2 -{\cal M}_4^2) 
\right. \nonumber\\
\left.
+\theta (x_3-x_1)  \,  v_{\rm inst \, 2}f_{\lambda}({\cal M}_{15}^2
 -{\cal 
M}_3^2)\, f_{\lambda}({\cal 
M}_{45}^2 -{\cal M}_2^2)\right\} \nonumber\\
f_{\lambda}({\cal M}_{12}^2 -{\cal M}_{34}^2) \Phi_{34} \label{div}
\end{eqnarray}

Assuming that the wave function $\Phi_{34}$ is bounded, it is
 straightforward to 
show that $I$ (\ref{conv}) (schematically, the 
integral containing  $(1-f^2)({\cal V}_{\rm inst} -{\cal V}_ 1-{\cal 
V}_2)$) is bounded and thus, has to be infrared convergent.

The remaining integral, $\rm I$ (\ref{div}) is  divergent as 
$x_5 \rightarrow 0$. 
This means that $x_3 \rightarrow x_1$ and $x_4 \rightarrow x_2$. Upon
 inspection of the
 integrand one can see that
the formfactors in (\ref{div}) restrict $k_5^{\perp}$ more and more
 severely 
as $x_5 \rightarrow 0$. It appears that, consequently, $\Phi_{34}
 \rightarrow \Phi_{12} $ 
as the divergence $x_5=0$ is approached.  So we add and subtract
 $\Phi_{12}$ to $\Phi_{34}$, i.e.
replace $\Phi_{34}$ in (\ref{div}) by 
$\left(\left[\Phi_{34}-\Phi_{12}\right]+\Phi_{12}\right)$. Then by
 expanding 
$\Phi_{34}$ in Fourier series around $\Phi_{12}$ it can be shown that
 the part 
of the integral with $\left[\Phi_{34}-\Phi_{12}\right]$ is, indeed,
 convergent as expected. 
Therefore, the infrared divergence is contained in 
\begin{eqnarray}
{\cal I} = - {g^2 \over{4 \pi^3}} C_F \int {\rm d}x_3 {\rm 
d}^2\kappa_{34}^{\perp} 
\left[ \theta (x_1-x_3) \, v_{\rm \, 1} f_{\lambda}({\cal M}_{35}^2
 -{\cal 
M}_1^2)\, f_{\lambda}({\cal M}_{25}^2 -{\cal M}_4^2) 
\right. \nonumber\\
\left.
+\theta (x_3-x_1)  \,  v_{\rm \, 2}f_{\lambda}({\cal M}_{15}^2 -{\cal 
M}_3^2)\, f_{\lambda}({\cal 
M}_{45}^2 -{\cal M}_2^2)\right] \nonumber\\
f_{\lambda}({\cal M}_{12}^2 -{\cal M}_{34}^2) \label{calI}
\end{eqnarray}
multiplying $\Phi_{12}$. This is similar to the divergent term from the
 one-body 
operator. 

To compare the divergence in the two-body operator with that of the
 one-body 
operator,  we need to isolate it.  
It is  sufficient to consider just series expansion for $x_5 
\rightarrow 0.$ Keeping leading and subleading terms,
arguments of the formfactors (\ref{arg}) reduce to
\begin{eqnarray}
{\cal M}_{25}^2 -{\cal M}_4^2 & = & {k_5^{\perp 2}\over{x_5}}\left(  x_2
  - 2 
x_5 { \kappa_{12}^{\perp} \cdot k_5^{\perp}\over{k_5^{\perp 2}}}+{\cal
 O}(x_5^2) 
\right)\\
{\cal M}_{35}^2 -{\cal M}_1^2 & = &  {k_5^{\perp 2}\over{x_5}}\left( 
{x_1^2\over{x_3}} + 2  x_5 {x_1\over{x_3}}{\kappa_{12}^{\perp} \cdot 
k_5^{\perp}\over{k_5^{\perp 2}}}+{\cal O}(x_5^2) \right)  \\
{\cal M}_{15}^2 -{\cal M}_3^2 & = &  {k_5^{\perp 2}\over{x_5}}\left( x_1
 - 2 
{\kappa_{12}^{\perp} \cdot k_5^{\perp}\over{k_5^{\perp 2}}}+{\cal
 O}(x_5^2) 
\right)  \\
{\cal M}_{45}^2 -{\cal M}_2^2 & = & {k_5^{\perp 2}\over{x_5}}\left( 
{x_2^2\over{x_4}}  + 2  {x_2\over{x_4}}{\kappa_{12}^{\perp} \cdot 
k_5^{\perp}\over{k_5^{\perp 2}}}+{\cal O}(x_5^2) \right) 
\end{eqnarray}
The overall similarity formfactor is in this limit
\begin{eqnarray}
f_{\lambda}( {\cal M}^2_{12} - {\cal M}^2_{34}) = 
e^{-{1\over{\lambda^4}}\, {[2 \kappa_{12}^{\perp} \cdot 
k_5^{\perp}+k_5^{\perp 2}]^2\over{x_1^2 x_2^2}} }\left[ 1+ {\cal
 O}(x_5)\right]
\end{eqnarray}

The formfactors contain factors $k_5^{\perp 2}/x_5$ which make the limit
 $x_5 
\rightarrow 0$ somewhat obscure. This can be evaded by a change of
 variables
$k_5^{\perp 2}/x_5 = {\cal K}^{\perp 2}$.
In these variables, 
\begin{eqnarray}
{\cal I} \propto \int {{\rm d}x_5 \over{x_5}} \left[\theta (x_1-x_3) 
r_{\delta}...
+ \theta(x_3 -x_1) r_{\delta}... 
\right] \int {\rm d}^2 {\cal K}^{\perp}  e^{-{(x_1^2
 +x_2^2)\over{\lambda^4}} 
{\cal K}^4} \left[1 + {\cal O}(\sqrt{x_5})\right] 
\end{eqnarray}
The subleading correction gives $\int {\rm d}x_5 (\sqrt{x_5})^{-1}$
 which is 
finite. The integral over ${\cal K}^{\perp} $ can be done analytically.
 Out of 
all infrared regulators, only those regulating $x_5\rightarrow 0$ are
 necessary, 
because the integral only diverges at its lower limit. The remaining
 infrared 
regulators can be set to 1. 
The result is
\begin{eqnarray}
- {g^2 \over{4 \pi^2}}C_F{\sqrt{\pi}\over{2}}\, {\lambda^2 
\over{\sqrt{x_1^2+x_2^2}}}\left[ 
\int_0^{1\over{x_1}} {{\rm d}y \over{y}}r_{\delta}\left( x_1 \, y\right)
r_{\delta}\left( x_2 \, y\right) +\int_0^{1\over{x_2}} {{\rm d}y 
\over{y}}r_{\delta}\left( x_1 \, y\right)
r_{\delta}\left( x_2 \, y\right) \right]  ,  \label{2bodydiv}
\end{eqnarray}
or,
\begin{eqnarray}
- {g^2 \over{4 \pi^2}}C_F{\sqrt{\pi}\over{2}}\, {\lambda^2 
\over{\sqrt{x_1^2+x_2^2}}}\left[ 
2 \int_0^{1} {{\rm d}y \over{y}}r_{\delta}\left( x_1 \, y\right)
r_{\delta}\left( x_2 \, y\right) + \log \left( {1\over{x_1 x_2}} \right)
 \right].
\end{eqnarray}
If this term multiplied by $\Phi_{12}$ is subtracted from the integral
 in 
(\ref{boundsteqn}), it becomes finite in the limit $\delta \rightarrow
 0$.

\subsection{Cancelation of IR divergences?}
Let us summarize the infrared divergent terms in the bound state
 equation
(\ref{boundsteqn}).

We have rewritten the bound state eqn. (\ref{boundsteqn}) as follows
\begin{eqnarray}
\lefteqn{\left( 4m^2 + 4m \, E\right) \Phi_{12}=}\nonumber\\
& &  {\kappa_{12}^2 +m^2(\lambda) \over{x_1 x_2}} \Phi_{12} 
\nonumber\\
& &  - {g^2 \over{4 \pi^2}} C_F \left\{
\int {{\rm d}x_3 {\rm d}^2\kappa_{34}^{\perp} \over{\pi}}
\left[ {\cal V}_1 + {\cal V}_2 + {\cal V}_{\rm inst}
 \right]f_{\lambda}({\cal 
M}_{12}^2 -{\cal M}_{34}^2) \Phi_{34} \right. \nonumber\\
 \lefteqn{ \hskip4truecm \left. - {\sqrt{\pi} \lambda^2 
\over{\sqrt{x_1^2+x_2^2}}} 
\int_0^{1} {{\rm d}y \over{y}}r_{\delta}\left( x_1 \, y\right)
r_{\delta}\left( x_2 \, y\right)  \Phi_{12} \right\} } \nonumber\\
  &+& {g^2\over{4 \pi^2}}\, 
C_F \, {\sqrt{\pi}\over{2}}\, \lambda^2 \left[ {1\over{\sqrt{2} \ x_1
 x_2}} 
 \int_0^1 \, {{\rm d}y\over{y}}\left(r_{\delta}(y)\right)^2   - 
 {2\over{\sqrt{x_1^2+x_2^2}}}
\int_0^{1} {{\rm d}y \over{y}}r_{\delta}\left( x_1 \, y\right)
r_{\delta}\left( x_2 \, y\right) 
\right]\Phi_{12} \label{boundsteqn2}
\end{eqnarray}
If the last line of (\ref{boundsteqn2}) vanished, the bound state
 equation would be
 independent of $\delta$, just as in previous formulation of the 
 similarity renormalization \cite{P2}, 
 or, as recently reported, in (2+1) dimensions \cite{haridch}.
 
It is obvious that this does not happen here, except for the leading
 order in the
 nonrelativistic expansion. Indeed, in the nonrelativistic limit
$x_{i} = 1/2$, and the two different functions of $x$es, constituting 
the coefficients of the infrared
 divergence from the one-body and from the two-body operators, have the
 same value
($4/{\sqrt{2}}$) at this point. However, an arbitrarily small deviation
 from $x=1/2$ 
introduces a positive divergent constant into the bound state equation
 in the 
limit $\delta \rightarrow 0$.
 The bound state equation therefore is not defined in this limit.

For any finite $\delta$, the bound state equation is well-defined and
 can be solved.
If $\delta$ is small, the
IR divergence forces the wavefunction to be peaked at $x=1/2$. Detailed
 behavior of 
the wavefunction is likely to depend on the specific form of
 $r_{\delta}$. In any case, 
the wavefunction width determines the allowed range of $x_5$.
 There is no such direct mechanism to restrict the transverse momenta,
 but they are 
 pushed to small values by the
formfactors in $\left[ {\cal V}_1 + {\cal V}_2 + {\cal V}_{\rm inst}
 \right]$ as 
$x_5$ is pushed to zero by $r_{\delta}$. It is peculiar that while the
 similarity formfactor
$f_{\lambda}({\cal  M}_{12}^2 -{\cal M}_{34}^2)$ also restricts the
 range of transverse
 momenta,
for $x_5 = 0$ this restriction is less severe than for any other
 possible $x_5$. This 
means that the implicit formfactors are more important when $x_5
 \rightarrow 0$.

If we assume that
\begin{eqnarray}
x_1 = 1/2 +\epsilon,
\end{eqnarray} 
with  $\vert \epsilon \vert << 1/2$, (consequently, $x_5<< 1/2$) and 
that other momenta, compared to mass, are also small, the
 bound state equation (\ref{boundsteqn2}) can be expanded. Care must be
 taken, however, 
 to keep track of the infrared divergence.
We find  
\begin{eqnarray}
\lefteqn{4 m \tilde{E} \, \Phi_{12}=}\nonumber\\
& &  4 \left[\kappa_{12}^2 + 4 \epsilon^2 m^2  + 
{m^2- m^2(\lambda)\over{m^2}} 
4 \epsilon^2 m^2 \right] \,
\Phi_{12} 
\nonumber\\
& -&   {g^2 \over{4 \pi^2}} C_F \left\{
\int {{\rm d}x_3 {\rm d}^2\kappa_{34}^{\perp} \over{\pi}} 4m^2
 f_{\lambda}({\cal 
M}_{12}^2 -{\cal M}_{34}^2) \Phi_{34}\right. \nonumber\\
& & \! \! \! \! \! \!
\left. \left[ {1\over{k_5^{\perp \, 2} +4 x_5^2 m^2}}\, \left\{ 1-
\exp\left(-  {(k_5^{\perp \, 2} +4 x_5^2 m^2)^2\over{2 x_5^2
 \lambda^4}}\right)
 \right\} %+ \right. \right.\nonumber\\
%& & \left. \left.
 + {1\over{ 4 x_5^2 m^2}}\,
\exp\left(-  {(k_5^{\perp \, 2} +4 x_5^2 m^2)^2\over{2 x_5^2
 \lambda^4}}\right)\,
r^2_{\delta}\left( {x_5\over{2}} \right)
 \right] \right. \nonumber\\
 \lefteqn{ \hskip4truecm \left. - {2\sqrt{\pi} \lambda^2 
\over{\sqrt{2}}} 
\int_0^{1} {{\rm d}y \over{y}}r^2_{\delta}\left( {y\over{2}}\right) 
 \Phi_{12} \right\} } 
\nonumber\\
  &+& {g^2\over{4 \pi^2}}\, 
C_F \, {3\sqrt{\pi}\over{\sqrt{2} }}\, \lambda^2  \, (2\epsilon)^2
 \int_0^1 \, {{\rm d}y\over{y}}r^2_{\delta}(y)  \Phi_{12}
 \label{NRboundsteqn}.
\end{eqnarray}
Here $\tilde{E}$ is the eigenvalue shifted by a constant from the one
 body operator,
 $\tilde{E} \equiv E - (m^2 - m^2(\lambda))/m$. 
On the right hand side of the equation, the third term is a correction
 to 
nonrelativistic kinetic energy. It is likely
 that this term is  small. The difference 
 between the current mass and
 mass including finite $\lambda$-dependent corrections can be expected
 to be negligible 
 for heavy quarks. 
 The  last expression in curly parenthesis is simply the subtraction of
 the IR divergence 
 in the two body operator to this order in momentum expansion, 
 so that the entire potential is final. The last line contains the 
 residual divergence for $x \neq 1/2$ 
 to its lowest nontrivial order, ${\cal O}\left( \epsilon^2\right)$ .
 
The structure of the right hand side becomes more clear upon 
 substitution
\begin{eqnarray}
2 x_5 m  =  q_z;  \  & k_5^{\perp} = q^{\perp} \\
2 \epsilon m =  - p_{ z};  \  & \kappa_{12}^{\perp} = p^{\perp} \\
\Phi_{12} \equiv \Phi (\vec{p} ); \   & \  \  \Phi_{34} \equiv  \Phi
 (\vec{p}+\vec{q}\, ) .
\end{eqnarray}
In these variables,
\begin{eqnarray} 
k_5^{\perp \, 2} +4 x_5^2 m^2 & = & q^2 \\
f_{\lambda}({\cal M}_{12}^2 -{\cal M}_{34}^2) 
& \simeq &  \exp \left( -{16\over{\lambda^4}} \left[2 \vec{p} \cdot
 \vec{q} +
 q^2\right]^2\right) .
\end{eqnarray}
The bound state equation (\ref{NRboundsteqn}) reads
\begin{eqnarray}
\lefteqn{4 m \tilde{E}  \Phi(\vec{p})= 4 \left[ p^2  
+ {m^2- m^2(\lambda)\over{m^2}} p_z^2\right] \Phi (\vec{p}) }
\nonumber\\
&-&   {g^2 \over{4 \pi^2}} C_F \left\{
\int {{\rm d}x_3 {\rm d}^2\kappa_{34}^{\perp} \over{\pi}} 4m^2 \right. 
 f_{\lambda}({\cal M}_{12}^2 -{\cal M}_{34}^2) \Phi(\vec{p}+\vec{q})
 \nonumber\\
& & \left. \left[ {1\over{q^2}}\, \left\{ 1-
\exp\left(-  {2 m^2 q^4\over{q_z^2 \lambda^4}}\right)
 \right\} + {1\over{ q_z^2}}\,
\exp \left(-  {2 m^2 q^4\over{q_z^2 \lambda^4}}\right)\,
r^2_{\delta}\left( {q_z\over{4m}} \right)
 \right]  \right. \nonumber\\
 \lefteqn{ \hskip4truecm \left. - {2\sqrt{\pi} \lambda^2 
\over{\sqrt{2}}} 
\left[\int_0^{1} {{\rm d}y \over{y}}r^2_{\delta}\left( {
 y\over{2}}\right)\right]
 \Phi (\vec{p}) \right\} } \nonumber\\
  &+& {g^2\over{4 \pi^2}}\, 
C_F \, {3\sqrt{\pi}\over{\sqrt{2} }}\, \lambda^2  \, {(p_z)^2\over{m^2}}
 \left[\int_0^1 \, {{\rm d}y\over{y}} r^2_{\delta}(y) \right]
 \Phi(\vec{p}) 
  \label{NRboundsteqn2}.
\end{eqnarray}.

The last line in (\ref{NRboundsteqn2}) is the uncanceled 
divergence which prevents us to take 
$\delta = 0$ limit at the Hamiltonian level.  Without numerically
 solving
 (\ref{NRboundsteqn2}) we cannot make any quantitative statements, 
 but some conclusions can be drawn just from the structure of the bound
 state equation. 
 If not for the Coulomb potential, i.e. the first integral term in 
 (\ref{NRboundsteqn2}),
it is obvious that with decreasing $\delta$, the wavefunction would 
get squeezed more and more. In the limit, the wavefunction would
 basically collapse to the point $x=1/2$ and  zero relative transverse
 momentum. This might 
allow  the eigenvalue to stay finite, and possibly roughly the same.
 With the Coulomb 
potential in place there could be some interplay between the positive
 $\delta$ divergence 
and the Coulomb. Note, however, that the Coulomb potential in 
our bound state equation is {\it incomplete} due to the presence of the
 exponential. 
The exponential cuts off the Coulomb potential at small $q$ (and nonzero
 $q_z$). 
It is still possible (although not guaranteed) that the mass of the
 eigenstate is finite 
and convergent in the $\delta \rightarrow 0$ limit, but at the expense
 of an  unphysical wavefunction.

\section{Summary and Conclusions}

Owing to nonzero mass of quark/antiquark, the only infrared divergence
 in the 
one- and two-body operators in a quark plus antiquark state is due 
to $x_5 \rightarrow 0$, i.e. the virtual gluon longitudinal fraction
 going to zero. 

The infrared divergent part of the  one-body operator obeys 
the one-body dispersion relation by construction of the similarity
 transformation. 
Specifically, it goes like a constant multiplying $x_i^{-1}$,  and is
 independent of the spectator. The expectation value in a
 quark/antiquark state is a simple
 sum of the quark and the antiquark contributions.
But the IR divergence from the two-body operator is a different, more 
complicated  function of $x$es. 
In addition to an overall explicit function, multiplying the divergent
 integral, whose form can
 be traced to the choice of $f_{\lambda}$, there is an implicit 
dependence through the arguments of the regulators. Thus a different
 $x$-dependence 
can be generated for different regulators, contrary to the one-body
 operator.
This occurs because boost invariance dictates the form of the one-body
 operator, and at the
 same time the two-body operator cannot lead to a function of the same
 form.
 
In the leading order nonrelativistic reduction,
all dependence on infrared regulator drops out. Beyond nonrelativistic
 leading order the
divergences do not cancel out in general, and it is not obvious that a
 regulator
such that the divergence would cancel out for all $x$, exists. Caution
 is in place when the
 nonrelativistic limit is considered, because the infrared divergence,
 even though related 
 to the most infrared divergent part of the effective terms, enters at
 the subleading order 
 in the momentum expansion.
 
 What are physical implications of this result and how to 
 evade the problem? This is a subject of an ongoing study. Here we only
 outline some
 possibilities. We wish to emphasize that at this point they are just
 speculations. 
 
 The first question one needs to address is whether the IR divergences
 are to be cured at the
  level of Hamiltonian. In previous section we argued that it is
 possible to solve for the bound
   state with any fixed value of $\delta$ and at the end look how the
 solution behaves in
    $\delta \rightarrow 0$. The eigenvalue can be expected to be finite
 and convergent,
     however, the wavefunction in the limit must collapse to a point.
 Such a wavefunction is
      counterintuitive. It is not necessarily a no-go, however.
 Wavefunction {\it by itself} is
       observable only through expectation values of other operators.
 These operators have to
        be regularized consistently with the Hamiltonian. 
        Then the behavior of observables need to be
         studied in the limit. 
         
This  approach was adopted in a recent work on QCD (2+1) by Chakrabarti
 and
 Harindranath \cite{haridch}. They solved the 
eigenvalue problem for a few formulations of similarity renormalization
 for Hamiltonian
 matrix elements. 
To avoid any confusion, we wish to emphasize that we use similarity
 renormalization for
 particles, and these two frameworks can have very different challenges.
 With this in mind, we note with interest that
 they find convergent eigenvalues, and some structure around $x=1/2$
 which they
  interprete as a sign of slow convergence.  
 
It may turn out that  removing the divergences at the level of
 Hamiltonian is necessary,
 or preferable. There are some compelling reasons for this. Leaving
 aside the open question
 of whether the limit of bound state solutions with finite $\delta$
 converges in the limit 
 $\delta \rightarrow 0$, improving our understanding (or lack of
 thereof) of the vicinity 
 of the zero mode could be very useful in constructing more convenient
 similarity
 transformations. In present work we used as our diagonal part in the
 similarity 
 transformation the current mass operator. Consequently, all formfactors
 are function
  of momenta compared to current mass. Even though in the bound state
 equation 
  the kinetic energy is expressed in terms of effective, or constituent,
 $\lambda$-dependent
  mass, any expansions in momenta have to be performed in reference to
 current mass.
  This is certainly not an issue in case of heavy quarks, but for light
 quarks it practically 
  rules out a simple nonrelativistic description of the bound state.
 It is not necessary, though, to use a $\lambda$-independent diagonal
 part to drive the
 similarity transformation. It could be devised around a
 $\lambda$-dependent, constituent
 mass, providing we have a resolution to the issue of  the IR
 divergences 
 at the Hamiltonian level.

One possibility is that understanding of the light-cone 
 zero mode is indeed crucial for finding the hadronic spectra, and that
 the requirement that
  the physical observables are in compliance with
Lorentz invariance is not sufficient to fully define the  IR cutoff
 theory and render it
 regularization independent. If this is the case, then we are in
 trouble. 

Another option that comes to mind is more down-to-earth. In a fully
 covariant theory, the
 physical state surely contains all Fock components. To maintain/restore
 symmetries
 order by order in coupling, it might be necessary to include in the
 state 
 at least those Fock components that can mix with $q\bar{q}$ at the
 given order. 
 To second order the only additional Fock component to include would be
 $q\bar{q} g$. It is
  possible that the infrared divergence in the $q\bar{q}$ cancels with
 an infrared divergence
   in 
$q\bar{q} g$. To support this view, recall, that in the standard
 perturbation theory the
 infrared divergence cancels out between instantaneous  exchange and one
 gluon
 exchange. Even though our
 framework differs from the usual perturbation theory, a similar
 cancelation might occur. The
  reason is that the formfactor structure of the mixing between
 $q\bar{q}$ and $q\bar{q} g$ is
 the same as that of the effective $q\bar{q}$ terms which are plagued by
 the infrared
 divergence. The main difference is that the mixing depends on a
 three-body wavefunction.
 Nevertheless, it is conceivable that both the  $q\bar{q}$ effective
 interaction and mixing with
  the three-body Fock component can be made finite by adding and
 subtracting the same IR
   counterterm, and it is possible that any ambiguities in connection
 with the choice of a
  specific IR regulator can be used to improve manifest rotational
 symmetry in the $q\bar{q}$ sector.

\section*{Acknowledgment}
This work was partially supported by DOE grant DE-FG02-97ER-41029, and
 by 
 DOE grant DE-FG02-87ER-40365. I would like to thank Cilka Marie and
 Stan G{\l}azek for discussions, Marek Karliner for useful suggestions,
 and
 Patty Halstead and Maciek Swat for technical assistance.

\end{document}